\documentclass[%
superscriptaddress,
 amsmath,amssymb,
 aps,
 notitlepage,
 twocolumn,
prb,
]{revtex4-2}

\usepackage{dcolumn}
\usepackage{multirow}
\usepackage{comment}
\usepackage{graphicx}
\usepackage[version=4]{mhchem}
\usepackage[colorlinks=True,linkcolor=red,citecolor=blue,urlcolor=blue]{hyperref}
\usepackage[utf8]{inputenc}
\usepackage[normalem]{ulem}
\newcommand{\bcvo}{\ce{BaCo2V2O8}}
\newcommand{\pcvo}{\ce{PbCo2V2O8}}
\newcommand{\scvo}{\ce{SrCo2V2O8}}
\graphicspath{{figs/}}

\begin{document}

\title{Magnetic structure and phase diagram of the Heisenberg-Ising spin chain antiferromagnetic PbCo$_{2}$V$_{2}$O$_{8}$}

\author{K.~Puzniak}
\affiliation{Helmholtz-Zentrum Berlin f\"{u}r Materialien und Energie GmbH, Hahn-Meitner Platz 1, D-14109 Berlin, Germany}
\affiliation{Institut f\"{u}r Festk\"{o}rperphysik, Technische Universit\"{a}t Berlin, Hardenbergstra{\ss}e 36, D-10623 Berlin, Germany}

\author{C. Aguilar-Maldonado}
\affiliation{Helmholtz-Zentrum Berlin f\"{u}r Materialien und Energie GmbH, Hahn-Meitner Platz 1, D-14109 Berlin, Germany}

\author{R.~Feyerherm}
\affiliation{Helmholtz-Zentrum Berlin f\"{u}r Materialien und Energie GmbH, Hahn-Meitner Platz 1, D-14109 Berlin, Germany}

\author{K.~Proke\v{s}}
\affiliation{Helmholtz-Zentrum Berlin f\"{u}r Materialien und Energie GmbH, Hahn-Meitner Platz 1, D-14109 Berlin, Germany}

\author{A. T. M. N. Islam}
\affiliation{Helmholtz-Zentrum Berlin f\"{u}r Materialien und Energie GmbH, Hahn-Meitner Platz 1, D-14109 Berlin, Germany}

\author{Y.~Skourski}
\affiliation{Dresden High Magnetic Field Laboratory (HLD-EMFL), Helmholtz-Zentrum Dresden-Rossendorf, D-01328 Dresden, Germany}

\author{L.~Keller}
\affiliation{Laboratory for Neutron Scattering and Imaging, Paul Scherrer Institut, Villigen CH-5232, Switzerland}

\author{B.~Lake}
\affiliation{Helmholtz-Zentrum Berlin f\"{u}r Materialien und Energie GmbH, Hahn-Meitner Platz 1, D-14109 Berlin, Germany}
\affiliation{Institut f\"{u}r Festk\"{o}rperphysik, Technische Universit\"{a}t Berlin, Hardenbergstra{\ss}e 36, D-10623 Berlin, Germany}

\begin{abstract}
The effective spin-1/2 antiferromagnetic Heisenberg-Ising chain materials, ACo$_2$V$_2$O$_8$, A = Sr, Ba, are a rich source of exotic fundamental phenomena and have been investigated for their model magnetic properties both in zero and non-zero magnetic fields. Here we investigate a new member of the family, namely PbCo$_2$V$_2$O$_8$. We synthesize powder and single crystal samples of PbCo$_2$V$_2$O$_8$ and determine its magnetic structure using neutron diffraction. Furthermore, the magnetic field/temperature phase diagrams for magnetic field applied along the {\bf c}, {\bf a}, and [110] crystallographic directions in the tetragonal unit cell are determined via magnetization and heat capacity measurements. A complex series of phases and quantum phase transitions are discovered that depend strongly on both the magnitude and direction of the field. Our results show that \pcvo is an effective spin-1/2 antiferromagnetic Heisenberg-Ising chain with properties that are in general comparable to those of \scvo and BaCo$_2$V$_2$O$_8$. One interesting departure from the results of these related compounds, is however, the discovery of a new field-induced phase for the field direction $H\|$[110] which has not been previously observed.

\end{abstract}

\date{\today}
\maketitle

\section{Introduction}

Quantum phase transitions (QPTs) have attracted considerable interest due to their relevance to the fundamental processes of quantum magnetism \cite{sachdev_2011}. Unlike a classical phase transition driven by thermal fluctuations, a QPT arises at $T$ = 0 K when the system is tuned by a non-thermal external parameter such as pressure, magnetic field, or chemical doping. The spin-1/2 spin-chain with Heisenberg-Ising (XXZ) exchange anisotropy, in a magnetic field applied transverse to the Ising direction generates one of the canonical examples of a QPT \cite{sachdev_2011}. The most famous experimental realization of this model was the quasi-one-dimensional (quasi-1D) spin-1/2 Ising ferromagnet CoNb$_2$O$_6$ \cite{Coldea_2010}. 
\par
More recently the quasi-1D antiferromagnetic materials AM$_2$V$_2$O$_8$, have been found to harbor a wealth of exotic phases including QPTs. Here the M-sites are filled by a magnetic transition metal ion such as  \ce{Cu^2+}, \ce{Ni^2+}, \ce{Co^2+} or \ce{Mn^2+}, while the divalent A-site ion and \ce{V^5+} are non-magnetic. Depending on the nature of the magnetic ion, different spin moments and anisotropies can be explored.
Of particular interest are the members ACo$_2$V$_2$O$_8$ where A = Sr, Ba, which give rise to effective 1D spin-1/2 antiferromagnets with Heisenberg-Ising (or XXZ) exchange anisotropy due to the \ce{Co^2+} ions which form 4-fold screw chains along the tetragonal {\bf c}-axis. The intrachain coupling is strong and antiferromagnetic, while the interchain coupling is weak and eventually gives rise to long-range antiferromagnetic N{\'e}el order at sufficiently low temperatures.
\par
In zero magnetic field, these compounds have a spinon continuum above the N{\'e}el temperature $T_N \approx 5$~K, and were used to demonstrate spinon confinement on cooling below $T_N$ where the continuum is replaced by sharp bound-spinon modes \cite{Grenier2015PRL,Wang2015,Bera2017}. A longitudinal magnetic field applied parallel to the easy axis which is the {\bf c}-axis also shows exotic physics. Above a critical field, the antiferromagnetic order is suppressed to much lower temperatures ($T<1$~K) and the systems undergo a transition to a longitudinal spin density wave and then a transverse canted antiferromagnet with increasing magnetic field \cite{ PhysRevB.87.054408, PhysRevB.92.134416,Shen_2019, Bera:2019zwb}. In the excitation spectrum, bound states of magnons known as Bethe strings, which were predicted by Hans Bethe in 1931 \cite{Bethe1931}, were observed for the first time in \scvo using terahertz spectroscopy \cite{Wang2018,PhysRevLett.123.067202} and inelastic neutron scattering \cite{Bera:2019zwb}. 
\par
The behavior of these chains in a transverse magnetic field along the {\bf a}-axis (perpendicular to the Ising anisotropy) is equally fascinating. Recent NMR measurements reveal two QPTs for \scvo as the magnetic order is suppressed by field \cite{ PhysRevLett.123.067203}. Neutron diffraction and inelastic neutron measurements for \bcvo revealed a quantum phase transition between two different types of solitonic topological objects \cite{Faure:2017iup} where the excitations can be described as collective solitonic modes superimposed on a continuum \cite{Faure2021}. At a lower magnetic field of 4.7~T within the ordered phase, a hidden 1D quantum phase transition was identified by NMR \cite{Zou2021}. It has universality class described by the exceptional $E_8$ Lie algebra which is characterized by eight gapped excitations whose gaps are theoretically predicted to have precise values \cite{ ZAMOLODCHIKOV1989}. These excitations were measured by inelastic neutron scattering \cite{Zou2021,wang2023spin} and terahertz spectroscopy \cite{Zhang2020} and compared successfully to theory \cite{ Wangx2021}. A magnetic field applied along the other transverse direction ([110]) was shown to give a  very different phase diagram than the {\bf a}-axis, with the N{\'e}el antiferromagnetic order found in zero field maintained to very high fields \cite{ Kimura2006,Kimura_2013, PhysRevB.87.224413, OKUTANI2015779}. The reason why the transverse [110]- and {\bf a}-field directions are different, is due to the complex g-tensor for the Co$^{2+}$ ions \cite{Kimura_2013,PhysRevB.87.224413} which is responsible for many of the unique properties of these magnets.
\par
The topic of this paper is a new and unexplored member of the ACo$_2$V$_2$O$_8$ family, namely PbCo$_2$V$_2$O$_8$. PbCo$_2$V$_2$O$_8$, with tetragonal space group $I4_1cd$ (\# 110) \cite{HE20071770}, is isostructural to \scvo and very similar to \bcvo (which has space group $I4_1/acd$ (\# 142)). As for \scvo and \bcvo the magnetic Co$^{2+}$ ions are arranged in edge-sharing CoO$_6$ octahedra forming 4-fold screw chains running along the {\bf c}-axis, which are well separated by non-magnetic V$^{5+}$ and Pb$^{2+}$ ions. There are four screw chains per unit cell, two rotating clockwise and the other two anticlockwise. Powder samples of \pcvo were synthesized previously \cite{HE20071770}, and magnetic and thermodynamic measurements reveal long-range N\'eel order below $T_{\mathrm{N}}\approx 4$~K \cite{HE20071770,HE2007404}, however, its magnetic structure has not been investigated. Under an external field, the powder sample shows a broad transition at $\mu_0H \approx 4$~T. 
\par
In this paper, we undertake the first detailed investigation of the magnetic properties of \pcvo. We synthesize powder samples and perform powder neutron diffraction to determine the magnetic structure. We also synthesize a large single crystal  which allows the anisotropic magnetism to be studied as a function of magnetic field applied along the {\bf c}, {\bf a}, and [110] directions. Using a combination of heat capacity and magnetization measurements we construct the magnetic field/temperature phase diagrams for these three directions. While the properties of \pcvo and rather similar to those of \scvo and \bcvo for the {\bf c}, {\bf a} axes, we discover a completely new phase for the field parallel to [110].




\begin{figure*}
	\centering
		\includegraphics[width=1.0\linewidth]{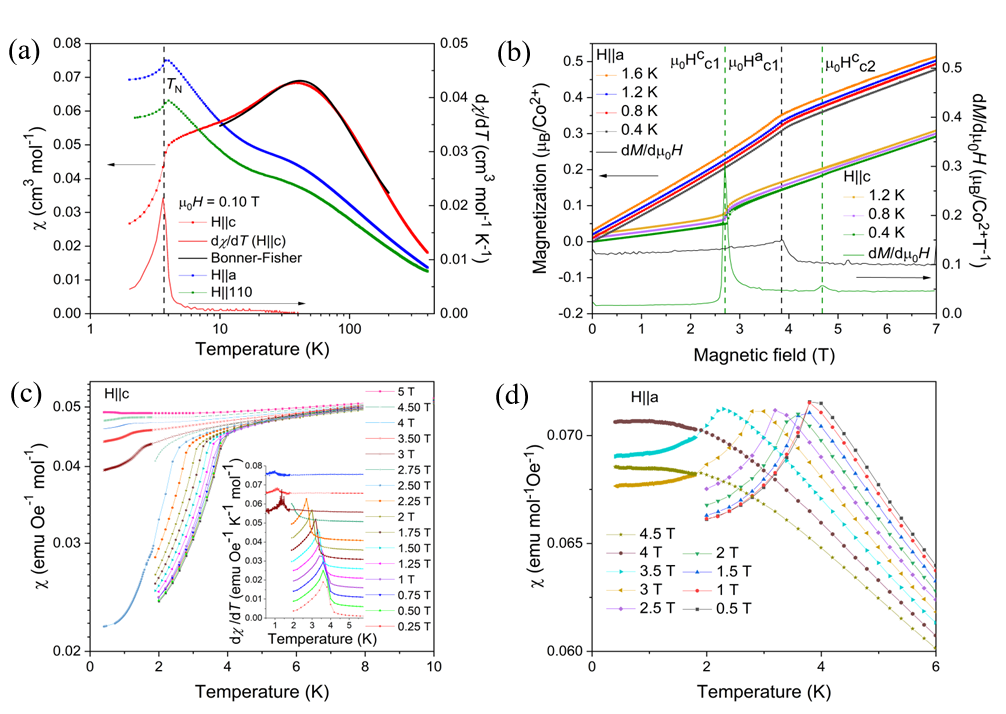}
	\caption{The DC magnetic susceptibility and the magnetization of \pcvo at low magnetic fields and temperatures. (a) Temperature-dependence of the susceptibility measured in a magnetic field of $\mu_0H = 0.10$~T applied parallel to the {\bf c}- ($\chi\|${\bf c}, red line), {\bf a}- ($\chi\|${\bf a}, blue line) and [110]- ($\chi\| [110]$, green line) directions. The solid red line gives the temperature derivative of $\chi\|${\bf c} and the vertical black dashed line indicates the N\'{e}el temperature, $T_{N}$ = 3.80 K. The solid black line show the fit to Equ. (1). (b) Field-dependence of the magnetization measured for $H\|${\bf c} at $T$ = 0.4, 0.8, and 1.2 K and also for $H\|${\bf a} at $T$ = 0.4, 0.8, 1.2, and 1.6 K. The green and black solid lines show d$M$/d$\mu_{0}$$H$ at $T = 0.4$~K for $H\|${\bf c} and $H\|${\bf a} respectively. (c) The low temperature susceptibility $\chi\|${\bf c}, measured for magnetic fields from $\mu_0H$ = 0.25 T to 5 T applied along the {\bf c}-axis. The inset shows the temperature dependence of the temperature derivative of susceptibility for magnetic fields from 0.25~T to 4~T. (d) Low temperature susceptibility $\chi\|${\bf a} measured for magnetic fields from $\mu_0H$ = 0.5 T to 4.5 T applied along the {\bf a}-axis.} 
	\label{fig:DCmagnsusc}
\end{figure*}

\section{\label{sec:ExpDet}Experimental Details}
Powder and single crystal samples of \pcvo were synthesized at the Core Lab Quantum Materials (CLQM), Helmholtz Zentrum Berlin f\"{u}r Materialien und Energie (HZB), Germany. The powder was prepared by the solid state reaction of high purity powders of PbO (99.99\%, Alfa Aesar), CoC$_2$O$_4$ $\cdot$ 2H$_2$O (99.995\%, Alfa Aesar), and V$_2$O$_5$ (99.99\%, Alfa Aesar) which were thoroughly mixed in the 1:1:2 molar ratio in ethanol and then sintered at 930$^{\circ}$ three to four times for 12 hours each with grindings performed after each sintering. For the crystal growth, a dense feed rod was prepared from the stoichiometric powder pressed under 2000 bars in a cold-isostatic-pressure (CIP) machine and subsequently sintered. Since \pcvo was found to melt incongruently, the traveling solvent floating zone technique was applied using a solvent excess in V$_2$O$_5$ at the tip of the feed rod. The Crystal growth was carried out in a 4-mirror type optical Floating Zone furnace (Crystal Systems Corp., FZ-T 10000-H-VI-VPO) with 150 W Tungsten halide lamps. It was performed in a 0.2 MPa Argon atmosphere at a growth rate between 0.5 to 1.0 mm/h. The as-grown single crystal was about 35 mm in length and about 5 mm in diameter. As far as we are aware these are the first reported crystals of this compound. 
\par
The crystal quality was checked using X-ray powder diffraction. A small piece of the crystal was crushed and ground into a powder. The powder diffraction pattern was collected at room temperature on a Bruker D8 diffractometer (Cu $K_{\alpha}$, energy 8.0478 keV, wavelength 1.5406 \AA). A long 2$\theta$ scan was done from 10 to 100 degrees with a step size of 0.0014 degrees coutning 7 seconds per point. X-ray Laue diffraction was also used to check the crystal quality and prepare oriented samples for thermodynamic and magnetic measurements.
\par
The field and temperature dependence of the magnetization were also measured at the CLQM, HZB. Measurements were performed using the Physical Properties Measurement System (PPMS 14 T Quantum Design) in magnetic fields up to $\mu_0H$ = 14 T over the temperature range from 1.8 K to 400 K, with field applied along the {\bf c}-, {\bf a}-, and [110]-axes. Measurements were also carried out using a Magnetic Property Measurement System (MPMS 7 T, Quantum Design), equipped with a \ce{^3He} insert in magnetic fields up to $\mu_0H$ = 7 T and over the temperature range from 0.4 K to 1.8 K for fields applied parallel to the {\bf c}- and {\bf a}-axes. High-field magnetization was measured at $T=1.5$~K in pulsed magnetic fields up to $\mu_0H$ = 58 T generated by the induction method using a coaxial pick-up coil system \cite{PhysRevB.83.214420} at the Hochfeld Magnetlabor Dresden in the Helmholtz-Zentrum Dresden Rossendorf (HZDR). The sample was cooled in zero field and when the desired temperature was stable, a magnetic field pulse of a total duration of 25~ms was applied. Measurements took place with field applied along the {\bf c}-, {\bf a}-, and [110]-axes. Normalization to absolute units was achieved by calibrating the data with the lower field PPMS magnetization obtained in static fields at the CLQM.
\par
The specific heat measurements were performed at the CLQM, by means of a relaxation method using the PPMS 14 T equipped with a \ce{^3He} insert. Magnetic fields up to 14 T were applied, and the temperature was varied between 0.4 K and 5 K (except in the case of zero field where the minimum temperature was 0.8 K because the low value of thermal coupling prevented lower temperature measurements). For each of the heat capacity scans, measured at different magnetic fields, an addenda measurement collected at 0 T was subtracted from the signal to obtain the sample heat capacity (note that the addenda of the used puck does not show any magnetic field dependence). Three crystal pieces were measured with the {\bf c}- (6.14 mg), {\bf a}- (5.10 mg), and [110]- (4.22 mg) axes respectively, parallel to the applied magnetic field. 
\par
Finally, the crystal and magnetic structure in zero magnetic field were investigated by neutron powder diffraction using the cold neutron diffractometer DMC, at the SINQ Facility in the Paul Scherrer Institute (PSI), Switzerland. The low-temperature measurements were performed using a \ce{^3He} stick inserted into an orange cryostat. A sample of mass $\simeq3$~g of the powder prepared by solid state reaction was sealed in a Copper can which was attached to the cold finger of the cryostat. Diffraction patterns were collected using a wavelength of $\lambda=2.46$~\AA\ for temperatures in the range 0.3~K to 120~K with typical counting times of six hours for 0.3~K($\ll T_{N}$), 4~K $(\approx T_{N}$) and 120~K, all other temperatures were counted for 2 hours. An additional high temperature measurement at 120~K was performed to study the structure of the sample, to reduce the background the \ce{^3He} insert was not used and the sample was loaded in a Vanadium can.

\section{Results}

\subsection{\label{sec:TdepMag}Temperature-dependence of the magnetization}

The static susceptibility and the temperature dependence of the magnetization of \pcvo was measured to explore its magnetic properties. Figure \ref{fig:DCmagnsusc}(a) shows the temperature dependence of the DC magnetic susceptibility of single crystal \pcvo measured in a field of $\mu_0H$ = 0.10 T applied parallel to {\bf c} ($\chi\|${\bf c}), {\bf a} ($\chi\|${\bf a}) and [110] ($\chi\| [110]$) in the temperature range from 2 K to 400 K. The temperature derivative of $\chi\|${\bf c} is also shown.

The $H\|${\bf c} data show a sudden drop below $\approx 4$~K and the temperature derivative of the susceptibility $d\chi\|${\bf c}$/dT$ shows a sharp peak at $3.80$~K indicating a transition to long-range antiferromagnetic order at $T_{N} = 3.80$~K. 
The magnetic susceptibilities $\chi\|${\bf a} and $\chi\| [110]$ show a peak and then a similar drop at $T_N$ which appears in their temperature derivatives as a peak (not shown) confirming this transition. While $\chi\|${\bf c} tends towards a very small value at low temperatures, $\chi\|${\bf a} and $\chi\| [110]$ tend to constant high values. This indicates that {\bf c}-axis is the easy axis or the Ising axis.
At higher temperatures a broad hump around 40~K for $\chi\|${\bf c} was observed which is a clear sign of short-range magnetic order probably due to the strong intrachain interactions expected in this compound which would give rise to quasi-one-dimensional behavior. The significant difference between $\chi\|${\bf c} compared to $\chi\|${\bf a} and $\chi\| [110]$, which persists even up to 400~K, is evidence for a large magnetic anisotropy (as was observed for \bcvo \cite{PhysRevB.72.172403} and \scvo \cite{OKUTANI2015779}). 

\begin{figure*}
	\centering
		\includegraphics[width=1.00\linewidth]{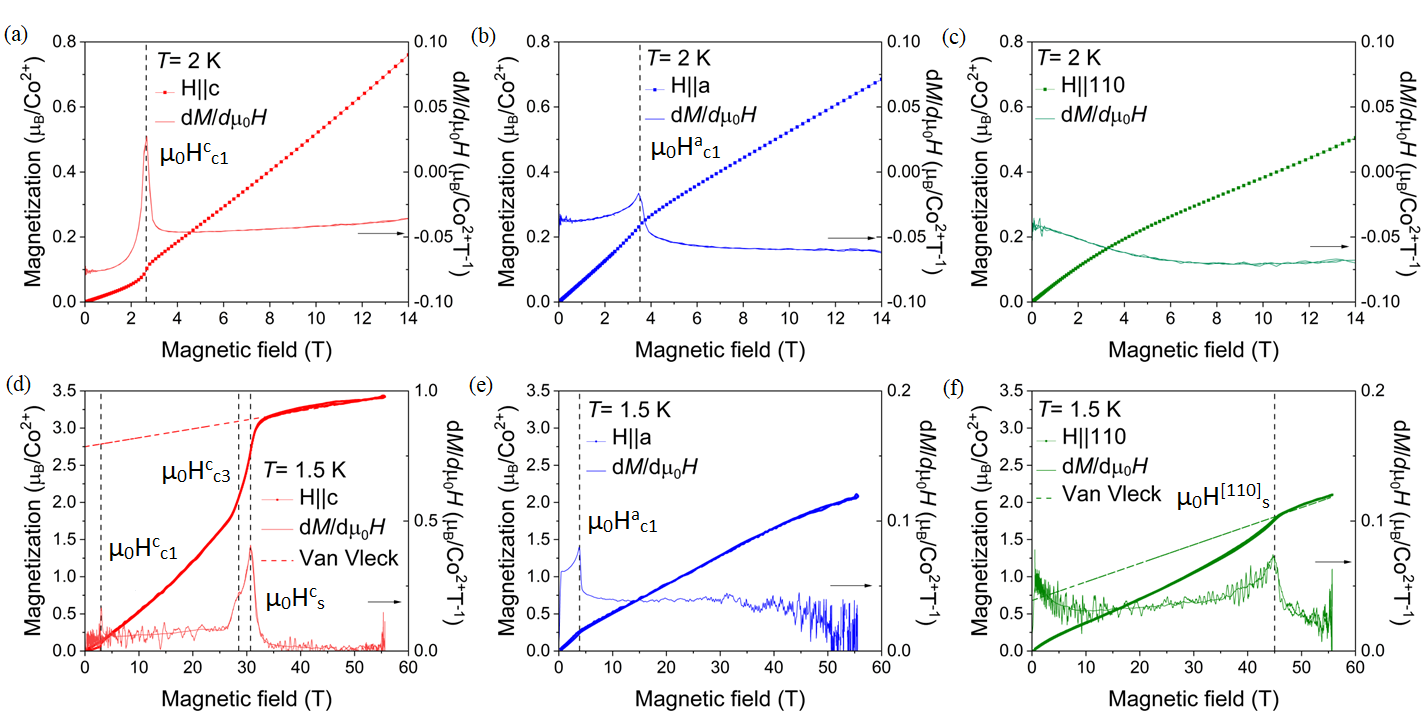}
		\caption{Field dependence of the magnetization of PbCo$_2$V$_2$O$_8$. The top panels show the magnetization curves measured at $T= 2$~K in the PPMS for magnetic fields up to 14 T and their field derivatives for (a) $H\|${\bf c} (red line), (b) $H\|${\bf a} (blue line), and (c) $H\|$[110] (green line). The lower panels give the high field magnetization curves collected using pulsed fields up to 58~T at $T$= 1.5 K, along with their field derivatives for (d) $H\|${\bf c} (red line), (e) $H\|${\bf a} (blue line) and (f) $H\|$[110] (green line). The vertical dashed black lines indicate the critical fields. The red dashed line in panel (d) and the green dashed line in panel (f) extrapolate the saturation magnetization to zero field and show the van Vleck contribution to the field dependence of the magnetization.}
  
  
	\label{fig:Magnetization}
\end{figure*}

Our susceptibility data are in general agreement with the previous powder susceptibility measurements for \pcvo which found a transition at $T_{N} = 4$~K \cite{HE20071770,HE2007404}. In order to estimate the intrachain interaction in PbCo$_2$V$_2$O$_8$, the $\chi\|${\bf c} susceptibility curve was fitted with the Bonner-Fisher model for uncoupled Ising chains \cite{PhysRev.135.A640, PhysRevLett.73.332, CHAKRABORTY20121945}:
\begin{eqnarray}
\chi (T) = \chi_{0}+\frac{N_A \mu_B^2 g_\parallel^2}{4k_B T}\frac{0.25+0.15x+0.30x^2}{1+1.98x+0.68x^2+6.06x^3},
\label{eq:ising-susc-para}
\end{eqnarray}
where $N_A$ is Avogadro's number, $k_B$ is the Boltzmann constant, and $\mu_B$ the Bohr magneton respectively. The term $g_\parallel$ is the Land{\'e} factor parallel to the Ising axis, $J$ is the intrachain exchange constant, $x$ = $J$/$k_B$$T$, and $\chi_{0}$ is a constant term. To avoid the effects of long-range magnetic order which are not included in this model, the fitted temperature range should start well above $T_{N}$. It should however include the characteristic broad hump at around 40~K which indicates the energy scale of the system. For the range 10 to 200~K, the fitted values of the parameters are $J = 32.10 \pm 0.06$~K and $g_\parallel = 5.30 \pm 0.01$ and $\chi_{0} = 0.0088 \pm 0.0001$ cm$^{3}$/mol, and the fitted curve is given by the solid black line in Fig.~\ref{fig:DCmagnsusc}(a). These values change a little if the lower limit is increased to 25~K which gives $J = 32.80 \pm 0.07$~K, $g_\parallel = 5.38 \pm 0.01$, and $\chi_{0} = 0.0075 \pm 0.0001$ cm$^{3}$/mol, revealing the reliability of the results and the applicability of the model.






Figure \ref{fig:DCmagnsusc}(c) shows the temperature dependence of the low-temperature DC magnetic susceptibility $\chi\|${\bf c} for several different magnetic field strengths from $\mu_0H = 0.25$~T to 5~T applied parallel to the {\bf c}-axis. For $0.25 < H\|${\bf c}~$< 2.7$~T, a strong decrease in susceptibility is observed at low temperatures. The temperature where the rapid drop occurs, which indicates $T_{N}(H\|${\bf c}), shifts with increasing magnetic field towards lower temperatures. It is more visible in the inset of the figure, as a peak in the temperature derivative of the magnetic susceptibility. This anomaly starts to disappear at 3~T and is gone above 4~T, although a low temperature transition is still visible as a small kink in the susceptibility and a very small maximum in the temperature derivative of $\chi\|${\bf c}. 

Figure \ref{fig:DCmagnsusc}(d) shows the low-temperature magnetic susceptibility $\chi\|${\bf a} at several different magnetic fields from $\mu_0H$= 0.5 T to 4.5 T applied along the {\bf a}-axis. The rapid drop in the magnetic susceptibility which indicates $T_{N}(H\|${\bf a}), shifts toward lower temperatures with increasing magnetic field and disappears completely at 4 T revealing a phase transition at around this field. Higher magnetic field measurements show no further transitions, suggesting the absence of a long-range magnetically ordered state.


\subsection{Field dependence of the magnetization}

We now investigate the magnetic field-dependence of the magnetization which provides information about the field-induced transitions. Figure \ref{fig:DCmagnsusc}(b) shows the low temperature magnetization up to 7~T for $H\|${\bf c} at $T = 0.4$, 0.8, and 1.2~K and also for $H\|${\bf a} at $T = 0.4$, 0.8, 1.2, and 1.6~K. The field derivative of the magnetization at $T = 0.4$~K for $H\|${\bf c} and $H\|${\bf a}  is also presented. For $H\|${\bf c} one can see two critical fields at $\mu_0H^{c}_{c1}$(0.4~K)~$= 2.70$~T and $\mu_0H^{c}_{c2}$(0.4~K)~$= 4.68$~T suggesting the presence of three distinct magnetic phases for fields up to 7~T at lowest temperatures. For $H\|${\bf a} there is one critical field at $\mu_0H^{a}_{c1}$(0.4~K)~$= 3.80$~T. These transitions appear almost independent of temperature in the studied temperature range.


Figures \ref{fig:Magnetization}(a), \ref{fig:Magnetization}(b), and \ref{fig:Magnetization}(c) show the magnetization curves along with their magnetic field derivatives at $T$ = 2 K as a function of applied magnetic field up to 14 T for $H\|${\bf c}, $H\|${\bf a}, and $H\|$[110], respectively. An abrupt increase in the magnetization is observed at $\mu_0H^{c}_{c1}$(2 K)~$= 2.65$~T for $H\| c$ which is clearly visible in d$M$/d$\mu_{0}$$H$ in Fig. \ref{fig:Magnetization}(a), indicating a field-induced transition. The second transition for this field direction, which was found at lower temperatures at $\mu_0H^{c}_{c2}$(0.4~K)~$ = 4.68$~T (see, Fig.~\ref{fig:DCmagnsusc}(b)), is not observed here at 2~K.
While no magnetization jump is seen for $H\|${\bf a} (see, Fig. \ref{fig:Magnetization}(b)) and $H\|$[110] (see, Fig. \ref{fig:Magnetization}(c)), a peak for $H\|${\bf a} is visible in the derivative d$M$/d$\mu_0H$ data, indicating the transition at $\mu_0H^{a}_{c1}$(2 K)~$ = 3.86$~T. 

We also measured the magnetization of \pcvo at much higher fields up to 58~T using a pulsed field magnet at $T = 1.5$~K. These data had to be calibrated, for which the DC magnetization curves up to 14 T at $T = 2$~K were used (Figs.~\ref{fig:Magnetization}(a), (b), and (c)). This approach is reasonable as the magnetization at 14~T does not change much with temperature in the range between 1.5 and 2~K. The normalized magnetization curves are shown along with their field derivatives for $H\|${\bf c} (Fig. \ref{fig:Magnetization}(d)), $H\|${\bf a} (Fig. \ref{fig:Magnetization}(e)) and $H\|$[110] (Fig. \ref{fig:Magnetization}(f)).

The magnetization curve for $H\|${\bf c} shows the first transition at $\mu_0H^{c}_{c1}$(1.5 K)~$ = 2.95$~T, while $\mu_0H^{c}_{c2}$ is not observable at this temperature. In the high field region, the magnetization is strongly nonlinear and appears to saturate at above $\approx 30$~T. This is confirmed by the field derivative of the magnetization. A peak is observed in d$M^c$/d$\mu_{0}H$ giving the saturation field along the {\bf c}-direction as $\mu_0H^{c}_s$(1.5 K)~$= 30.7$~T. This peak has a shoulder indicating a third transition at the slightly lower field of $\mu_0H^{c}_{c3}$(1.5 K)~$= 28$~T. Above the saturation field we can see a linear increase of magnetization. This increase is related to the van Vleck contribution to the magnetization which is linear with field. The van Vleck contribution was fitted and extrapolated to zero magnetic field (red dashed line). It is estimated to be $\chi_{VV}$ = 0.012 $\mu_{B}$/T per Co$^{2+}$ thus giving the saturation value of the magnetization as $M^c_s= 2.75$~$\mu_{B}$. Assuming effective spin$-1/2$ moments on the Co$^{2+}$ ions, this value of $M^c_s$ suggest that $g_{\parallel} = 5.5$ in agreement with the value found from static susceptibility. 


The magnetization curve for \pcvo with $H\|${\bf c} is similar to that found previously for \scvo \cite{OKUTANI2015779} and \bcvo \cite{Kimura2006,PhysRevLett.123.067202}. For \scvo the saturation field is 28.3~T 
and the saturation magnetization is $3$~$\mu_{B}$ \cite{OKUTANI2015779}. A high field transition just below the saturation at 23.7~T was also observed  \cite{OKUTANI2015779} comparable to the transition at $\mu_0H^{c}_{c3}$(1.5 K)~$= 28$~T found in PbCo$_2$V$_2$O$_8$. For \bcvo the critical fields are at 19.5~T (transition) and 22.7~T (saturation), and the saturation magnetization is $2.5-3.2$~$\mu_{B}$ \cite{Kimura2006,kimura2008,PhysRevLett.123.067202}.

The high field magnetization curve for $H\|${\bf a} shows the first transition as a change of slope and as a peak in d$M^a$/d$\mu_{0}H$ occurring at $\mu_0H^{a}_{c1}$(1.5~K)~$ = 3.52$~T. The magnetization then increases approximately linearly up to 40~T. Above 45~T a slight rounding of the magnetization curve is observed. A possible tendency towards saturation of the high field magnetization was also reported for \bcvo and \scvo \cite{Kimura_2013,OKUTANI2015779}.

The magnetization curve for $H\|$[110] is approximately linear and does not show any transitions at 1.5~K up to 40~T. 
Above 40~T its slope increases and at 45~T it becomes flatter suggesting saturation at $\mu_0H^{[110]}_s$(1.5~K)~$= 45$~T as indicated by the field derivative. Similarly as for $H\|${\bf c} direction, the van Vleck contribution for $H\|$[110] direction was fitted and extrapolated to zero magnetic field. It is estimated to be $\chi_{VV}$ = 0.025 $\mu_{B}$/T per Co$^{2+}$ thus giving the saturation value of the magnetization as $M^{[110]}_{s}= 0.68$~$\mu_{B}$. Comparable behavior was observed for \scvo  \cite{OKUTANI2015779}, for which the saturation field is  $\mu_0H^{[110]}_s$(1.4 K)~$ = 45.7$~T. An additional high field transition, not observed in PbCo$_2$V$_2$O$_8$, was found at 33.0~T. For \bcvo saturation is observed at 40.9~T with a value of $\approx 1.35$~$\mu_{B}$ and the additional transition is found at 30.8~T \cite{Kimura2006,Kimura_2013,PhysRevB.87.224413}. 



By analogy to \bcvo and \scvo the difference between the magnetization curves for $H\|${\bf a} and $H\|$[110] for \pcvo may be explained by the 4-fold screw chain structure of the CoO$_6$ octahedra. These octahedra are distorted and their apical bond is tilted away from the {\bf c}-axis by a few degrees in a direction that follows the screw chain rotation. These features give rise to a complicated g-tensor in both \bcvo~\cite{Kimura_2013} and \scvo~\cite{PhysRevLett.123.067203}. As a result, a magnetic field applied along the {\bf a}-axis gives rise to an effective field parallel to the {\bf b}-axis, which is staggered along the chain driving the spin-flop transition of the spins from pointing along the {\bf c} to the {\bf b} direction at $\mu_0H^{a}_{c1}$~\cite{Kimura_2013, Faure:2017iup}. In contrast, no such staggered field occurs when the field is applied in $H\|$[110]-direction and therefore no spin-flop transition is observed. It should be noted that our diffraction results described in Section \ref{sec:MagStruct} find that \pcvo  also has the moments canted by a small angle away from the {\bf c}-axis implying that a similar mechanism could apply here. 

\subsection{\label{sec:HC}Heat capacity}

\begin{figure}[h!]
\includegraphics[width=1.0\linewidth]{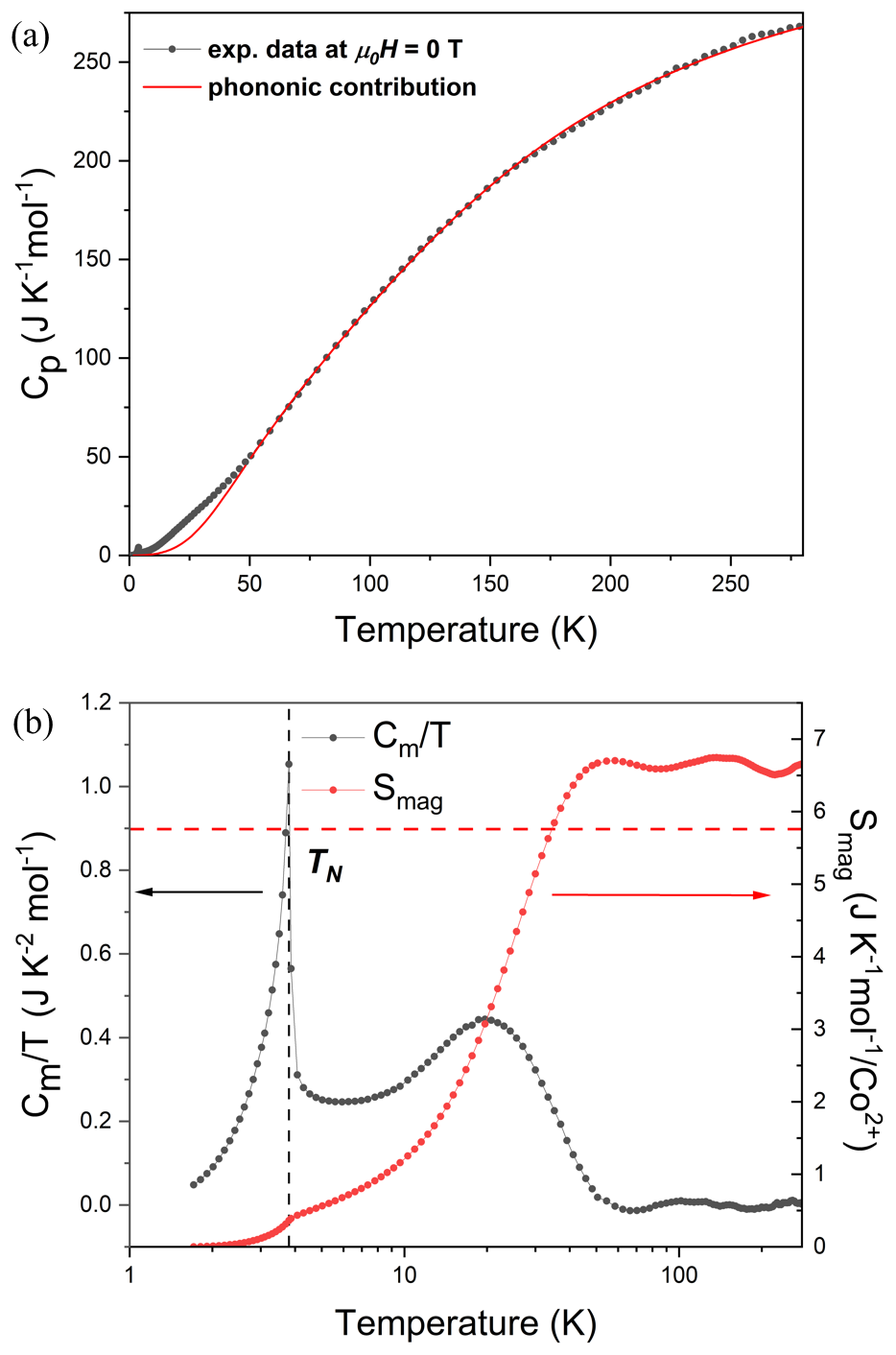}
\caption{\label{fig1}\ (a) Temperature-dependence of the heat capacity measured in zero field from 1.7 K to 280 K (black dots). The data above 50 K which is dominated by the lattice contribution were fitted to a combination of Einstein and Debye terms (red line). (b) The temperature-dependence of the magnetic heat capacity divided by temperature $C_m(\mu_0 H=0$~T$)/T$, obtained after subtraction of the phononic contribution, as well as the magnetic entropy $S_{mag}$ at $\mu_0H = 0$~T are shown by the black and red dots respectively. The horizontal red dashed line marks the value $S_{mag}$ = R ln(2), expected for the effective spin-1/2 magnetic moments.}
\label{fig:zerofield}
\end{figure}

\pcvo was also investigated using heat capacity measurements which provide a very accurate way to identify the phase transitions. Both temperature- and field-dependent measurements were performed in zero magnetic field and with the field applied along the $\mu_0H\|${\bf c}, $\mu_0H\|${\bf a} and $\mu_0H\|$[110] directions. Figure \ref{fig:zerofield}(a) shows the temperature dependence of the heat capacity $C_p$ from $T = 280$~K down to $T = 1.7$~K measured in zero field. The data above 50 K up to 280 K where the magnetic contribution becomes negligible, were fitted by Einstein and Debye terms to model the phononic contribution which was then extrapolated down to base temperature as shown by the solid red line. The magnetic heat capacity $C_m(\mu_0H=0$~T$)$, was extracted by subtracting this contribution. $C_m(\mu_0H=0$~T$)/T$ is presented in Fig.~\ref{fig:zerofield}(b) by the black dots and shows a sharp $\lambda$-type anomaly at $T_N (\mu_0H=0$~T$) = 3.80$~K. Finally, we perform the integral over temperature of the magnetic heat capacity divided by temperature in order to obtain the magnetic entropy as shown by the red dots. The magnetic entropy saturates above 40~K at $\approx$~6.65~JK$^{-1}$ per mole of Co$^{2+}$, a value slightly larger than the value $S_{mag}$ = R ln(2) = 5.76~JK$^{-1}$ per mole of Co$^{2+}$ where R is the gas constant, expected for a doublet ground state. This result suggests that we can assign an effective spin-1/2 moment to the Co$^{2+}$ ions.

\begin{figure*}
\includegraphics[width=1\textwidth]{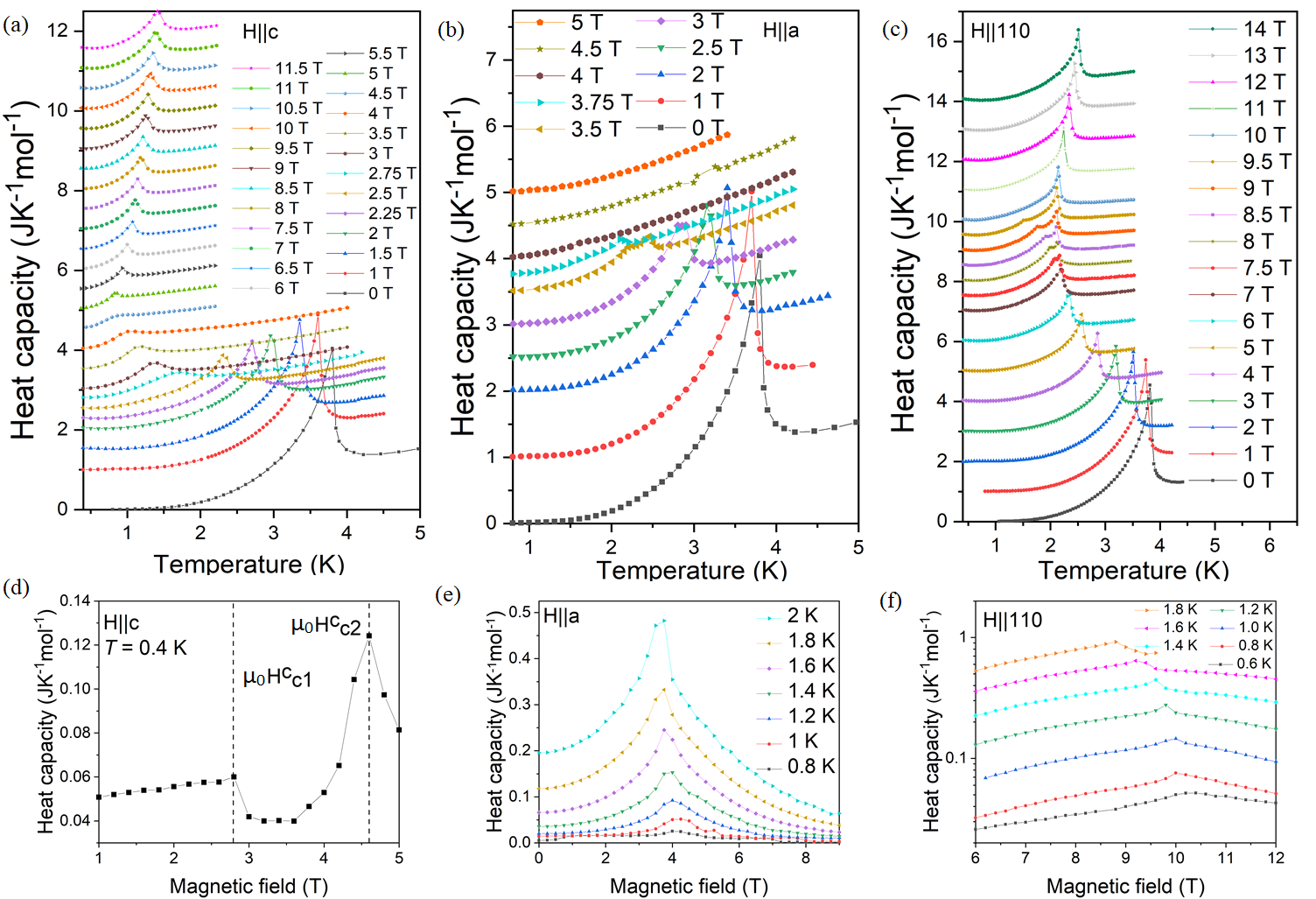}
\caption{\label{fig1}\ Heat capacity of single crystalline PbCo$_2$V$_2$O$_8$. Temperature scans $C_p$($T$) were performed under various values of magnetic field applied (a) longitudinally $H\|${\bf c}, (b) transversely $H\|${\bf a}, and (c) transversely $H\| [110]$. Panels (d), (e), and (f) show field scans of $C_p$($H$) for $H\|${\bf c}, $H\|${\bf a}, and $H\| [110]$ respectively, at various temperatures. Heat capacity data at non-zero field were collected in the temperature range above 0.4 K; however, for zero field the data were collected starting from 0.8 K as explained in Section~\ref{sec:ExpDet}. An offset proportional to an applied magnetic field with a proportionality factor of 1 JK$^{-1}$mol$^{-1}$/T has been added to each heat capacity curve in panels (a), (b), and (c) for clarity. In the case of panels (d), (e), and (f) no offset has been added. The heat capacity data are presented as raw data without subtraction of the phononic contribution which is very small at these low temperatures.}
\label{fig:fig1}
\end{figure*}

Figure \ref{fig:fig1}(a) shows the temperature-dependence of the heat capacity at low temperatures for a single crystal of \pcvo under longitudinal magnetic field $\mu_0H\|${\bf c} from $\mu_0H = 0$~T to 11.5~T. The measurements started from 0.4~K except for zero field where the lowest temperature was 0.8~K as explained in Section~\ref{sec:ExpDet}. At $\mu_0H = 0$~T the heat capacity curve shows a sharp $\lambda$-type anomaly at $T_N (\mu_0H=0$~T$) = 3.80$~K, indicative of the second-order phase transition between the paramagnetic and N\'{e}el phases. These data are in general agreement with the magnetic heat capacity divided by temperature presented in Fig. \ref{fig:zerofield}(b) which was collected on another sample, and with previous measurements on a powder sample of \pcvo which found the transition at $T_{N} = 4$~K \cite{HE20071770,HE2007404}. Below the transition, the heat capacity decreases very rapidly with decreasing temperature suggesting that the magnetic excitations are gapped as would be expected if there were Ising anisotropy. To estimate the gap size, a Schottky term was fitted to the zero field data in the range from 0.8~K to 2~K (well below the transition) and yielded the energy gap $\Delta E = 0.94 \pm 0.04$~meV. 
\par
With increasing longitudinal magnetic field ($H\|${\bf c}), the peak shifts to lower temperatures and decreases in amplitude rapidly. It almost disappears at $\mu_0H^{c}_{c1} \approx 2.75$~T which is close to the value of $\mu_0H^{c}_{c1}$(0.4~K)~$= 2.70$~T found from magnetization.
The peak reappears at 3~T at 1.25~K suggesting the appearance of a new phase at $\mu_0H > 3$~T. It becomes weak and broad again at $\mu_0H^{c}_{c2} \approx 4.5$~T implying the second phase boundary.
Interestingly at 5~T, the peak reappears again and shifts to higher temperatures with increasing field reaching 1.4~K at 11.5~T. This peak marks the upper temperature boundary of a high field phase, which is discussed in the next section.
The two field-induced phase transitions can be seen more clearly in the field-dependence of the heat capacity for $H\|${\bf c} at $T=0.4$~K presented in Fig. \ref{fig:fig1}(d). Two anomalies are observed giving the critical fields $\mu_0H^{c}_{c1}(T=0.4$~K$) = 2.79$~T and $\mu_0H^{c}_{c2}(T=0.4$~K$) = 4.60$~T in good agreement with the low temperature magnetisation results.
\par
Figure~\ref{fig:fig1}(b) shows the temperature dependence of the heat capacity under transverse field $\mu_0H\|${\bf a}, from $\mu_0H = 0$~T to 5~T. The amplitude of the $\lambda$ anomaly gradually decreases as the magnetic field is increased, and the peak shifts to lower temperature, finally disappearing at around $\mu_0H^{a}_{c1} \approx 4.0$~T. No further anomalies were observed at higher fields. This result is consistent with magnetization data for $\mu_0H\|${\bf a} (see, Fig. \ref{fig:DCmagnsusc}(d)). The field-dependence of the heat capacity for $\mu_0H\|${\bf a} for temperatures from 0.8~K to 2.0~K is presented in Fig.~\ref{fig:fig1}(e). A peak is found whose amplitude decreases with decreasing temperature and it is very weak at  0.8~K. The position of the peak is almost independent of temperature and indicates the phase transition at $\mu_0H^{a}_{c1}$(0.8~K)~$\approx 4.0$~T. 
\par
For the case of $\mu_0H\| [110]$ the temperature-dependence of the heat capacity was measured for various magnetic field values up to $\mu_0H = 14$~T (see Fig.~\ref{fig:fig1}(c)). The critical temperature and the amplitude of the $\lambda$-anomaly decrease as $\mu_0H$ increases up to 7~T. Starting from 7~T we observe a subtle splitting of the peak, which increases up to 9~T. For $\mu_0H > 10$~T, the splitting vanishes. To follow this feature more carefully, the heat capacity was scanned over the field range from 6~T to 12~T for several temperatures from 0.6~K to 1.8~K (see, Fig. \ref{fig:fig1}(f)). At 1.8~K, we observe a cusp at 
$\mu_0H^{[110]}_{c1}$(1.8~K)~$\approx 8.5$~T that moves to higher fields with decreasing temperature. This marks the start of a new phase boundary which can be traced to $\mu_0H^{[110]}_{c1}$(0.6~K)~$\approx 10.4$~T. This critical field was not observed in our DC magnetization because this was measured at 2~K.
\subsection{\label{sec:MPD}Magnetic phase diagram}

\begin{figure}[h!]
	\centering
	\includegraphics[width=1.00\linewidth]{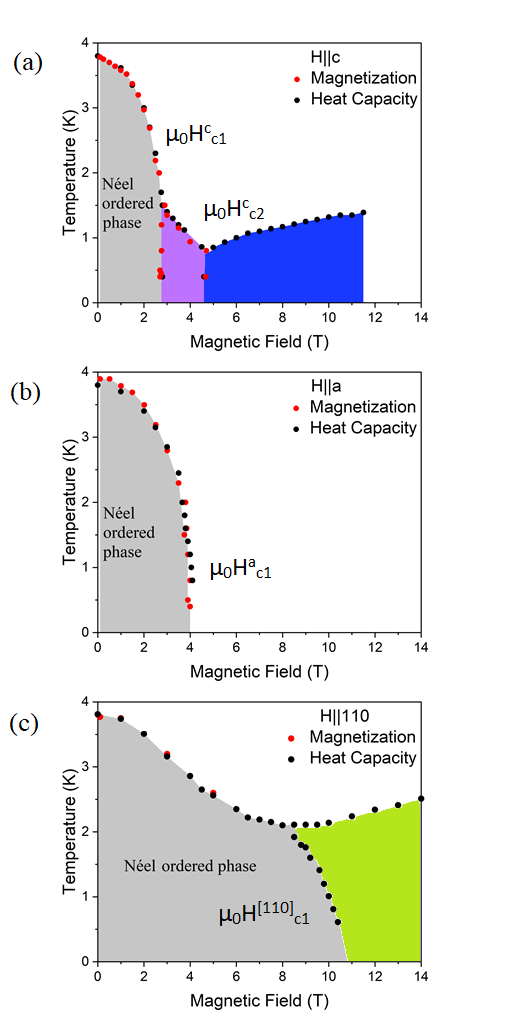}
	\caption{Phase diagram of \pcvo as a function of magnetic field and temperature for (a) longitudinal field $H\|${\bf c}, (b)  transverse field $H\|${\bf a}, and (c) transverse field $H\|$[110]. The black circles represent the critical fields and temperatures determined by the heat capacity measurements and the red circles represent the critical fields and temperatures extracted from the magnetization measurements. The different colored regions indicate the different magnetic phases. The phase colored gray is identified as a N\'{e}el antiferromagnetic phase in \ref{sec:MagStruct}. By analogy with \scvo and \bcvo the phase colored pink might have longitudinal spin density wave order, whereas the phase colored blue might be a transverse antiferromagnetic phase, although these field-induced phases have yet to be confirmed by neutron diffraction.} 
	\label{fig:Phase}
\end{figure}

The magnetic and heat capacity measurements were used to extract the magnetic field/temperature phase diagrams of \pcvo which are presented in Fig.~\ref{fig:Phase} for the field directions $\mu_0H\|${\bf c}, $\mu_0H\|${\bf a}, and $\mu_0H\|$[110]. The phase boundaries are obtained from the  positions of the extremes in the field derivatives of $M(H)$ and temperature derivative of $\chi(T)$ (red circles) and from the positions of the peaks in $C_p$($T$) and $C_p$($H$) (black circles). 

Figure \ref{fig:Phase}(a) shows the phase diagram for magnetic field applied along the {\bf c}-axis where three magnetic phases are present up to 11.5 T. The boundaries between these phases appear to extend down to zero Kelvin. In order to estimate the critical fields at absolute zero, the first phase boundary was fitted with the empirical power law formula $T_{c}=A(H_{c}-H)^{\phi}$ over the temperature range from 2 to 2.6 T, \cite{PhysRevLett.123.067203} which yielded the critical field value of $\mu_0H^{c}_{c1}$(0~K)~$ = 2.73 \pm 0.02$~T and exponent $\phi=0.27 \pm 0.01$. The extrapolation of the second phase boundary gives the value of the second critical field $\mu_0H^{c}_{c2}$(0~K)~$= 4.66$~T. 
This phase diagram resembles the $H\|${\bf c} phase diagrams found for \scvo \cite{ PhysRevB.105.174428, Shen_2019,Bera:2019zwb} and \bcvo 
\cite{PhysRevB.72.172403,PhysRevB.92.060408,PhysRevB.87.224413,kimura2008,PhysRevB.92.134416}, where three magnetic phases were also found.
These phases were identified, using neutron diffraction, as N\'{e}el antiferromagnetic (low fields), longitudinal spin density wave (intermediate fields), and transverse antiferromagnet (high fields) \cite{PhysRevB.83.064421,PhysRevB.87.054408, Shen_2019,PhysRevB.92.134416}. As for PbCo$_2$V$_2$O$_8$, the second and third phases have significantly lower ordering temperatures compared to the N\'{e}el phase. Just above these ordering temperatures a quantum critical regime is predicted. While qualitatively similar to \scvo and BaCo$_2$V$_2$O$_8$, \pcvo appears to have a smaller energy scale with the transitions occurring at somewhat lower fields.

Figure~\ref{fig:Phase}(b) shows the magnetic phase diagram of \pcvo for field applied along the {\bf a}-axis. Only one phase boundary is visible in this case. This boundary was fitted with the power law function over the temperature range from 3 to 4 T and yielded the value of critical field at $T=0$~K of $\mu_0H^{a}_{c1}$(0~K)~$= 4.01 \pm 0.02$~T and field exponent $\phi$ = $0.30 \pm 0.03$. The $H\|${\bf a} phase diagrams for \scvo and \bcvo are similar with a single phase that terminates at the critical fields of 7~T \cite{PhysRevLett.123.067203,PhysRevB.103.144405} and $\approx 10$~T \cite{PhysRevB.87.224413,Faure:2017iup,Kimura_2013} respectively. This transition was described as a quantum phase transition from N\'{e}el order to a quantum disorder regime \cite{PhysRevLett.123.067203} or a topological quantum phase transition between two different types of solitonic topological objects \cite{Faure:2017iup}.

Figure \ref{fig:Phase}(c) shows the magnetic phase diagram of \pcvo with the uniform magnetic field applied along the $H\| [110]$ direction. The critical temperature decreases smoothly with increasing field up to 8~T. Starting from 8~T one can clearly see a phase boundary to a new ordered phase, whose critical field increases to 10.4~T as temperature is reduced to 0.6~K. By extrapolation, we  estimate that this new phase boundary may persist up to $\mu_0H^{[110]}_{c1}$(0~K)~$= 10.9$~T. Interestingly, the transition temperature increases above 8~T as this new phases emerges. This phase was not  observed in the two other sister compounds \scvo or \bcvo where no phase transitions were observed below 30~T. However, a transition is observed in \scvo at the much higher field of 33.0~T \cite{OKUTANI2015779} and in \bcvo at $\approx 30$~T \cite{Kimura2006,Kimura_2013,PhysRevB.87.224413}. The nature of this transition and the high field phase have never been explored.
It is not clear whether this transition is related to the $\mu_0H^{[110]}_{c1}$(0 K) = 10.9~T transition in \pcvo which occurs at much lower fields. In any case, the relatively low critical field of the transition makes it accessible to experimental techniques such as neutron scattering, providing an opportunity for its exploration.

\subsection{\label{sec:MagStruct}Magnetic structure at zero magnetic field}

\begin{figure}[!ht]
	\centering
	\includegraphics[width=1.0\linewidth]{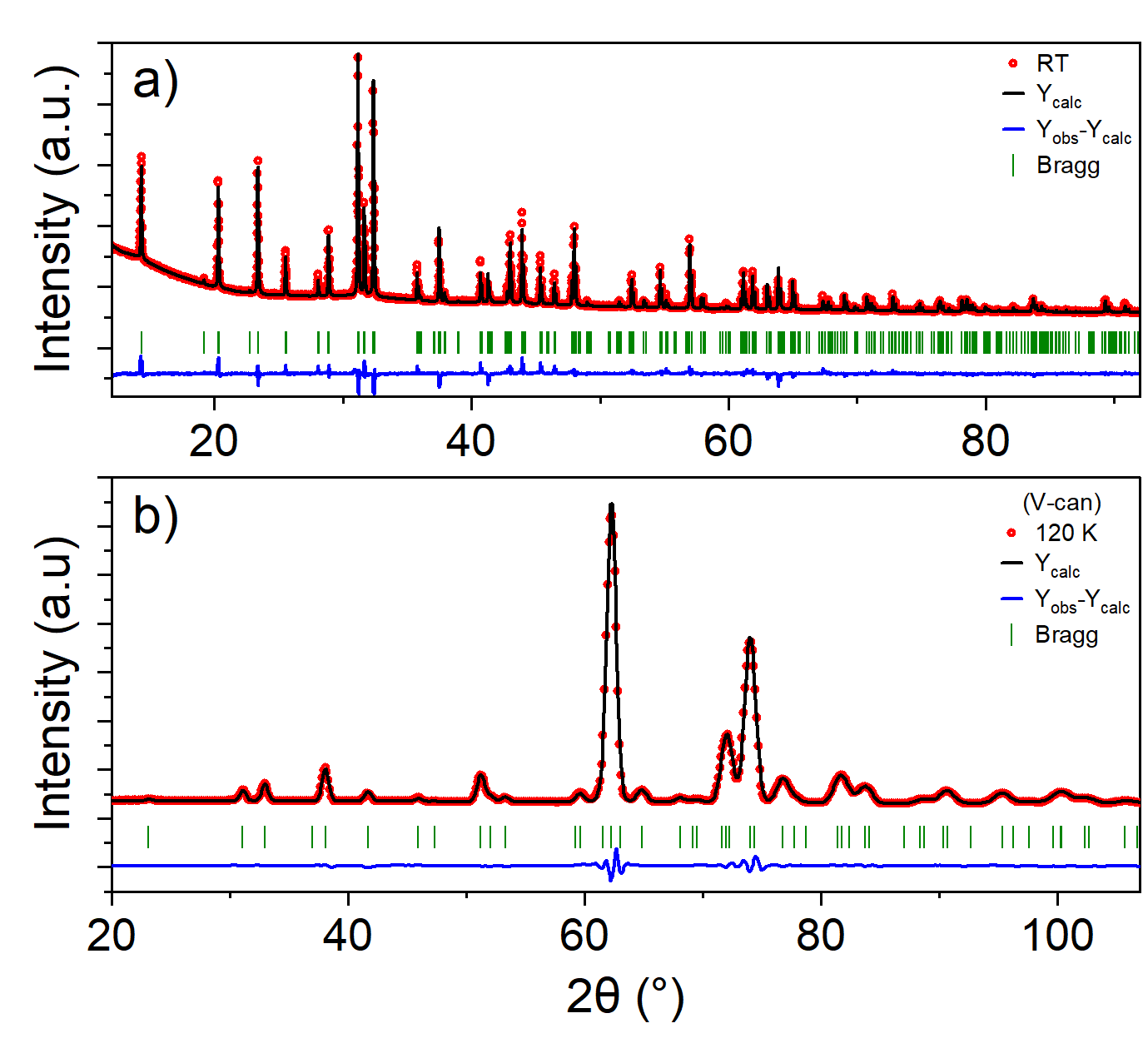}
	\caption{Powder patterns of \pcvo collected at high temperatures in zero magnetic field showing (a) X-ray powder diffractogram for the crushed single crystal at room temperature, and (b) Neutron powder diffraction pattern from the powder sample at $T$ = 120 K. In both cases the observed and calculated intensities are represented by the open red circles and solid black line respectively. The difference between them is given by the blue solid line and the Bragg peak positions for the $I4_{1}cd$ space group are given by the vertical green lines.}
	\label{fig:RietvStruct}  
\end{figure}

This section focuses on the magnetic structure in the low temperature and low magnetic field phase of \pcvo by analysing neutron powder diffraction data collected in zero magnetic field. However we first verify the crystal structure and the quality of the sample using X-ray and neutron powder diffraction experiments at high temperatures. The X-ray powder diffraction data was obtained from a crushed single crystal at room temperature (see Fig.~\ref{fig:RietvStruct}(a)), meanwhile the neutron powder diffraction was performed on the powder sample at $T = 120$~K that was measured without the \ce{^3He} insert to avoid extra background peaks from the equipment (see Fig.~\ref{fig:RietvStruct}(b)). The two methods are complementary, since unlike X-ray, neutrons are sensitive to light elements like Oxygen. On the other hand, Vanadium, which does not contribute to the neutron Bragg peaks because it only scatters incoherently, is observable by X-rays.   
\par
\begin{table}[]
\caption{Atomic coordinates and isotropic Debye-Waller factors obtained from the combined refinement of the neutron powder diffraction pattern of the crushed single crystal at $T = 120$~K (using DMC, PSI) and the X-ray powder pattern at $T = 300$~K from the powder sample (using Bruker D8, HZB). The refinement was performed in space group $I4_1cd$ (\# $110$) yielding lattice parameters $a=b=12.3580(8)$~{\AA}, $c= 8.4444(5)$~{\AA} at room temperature with $R_{F}= 10.5$\% and $R_{wp}=13.5$\% from the X-ray data, and $a=b=12.2610(12)$~{\AA}, $c= 8.3741(10)$~{\AA} at 120~K with $R_{F}= 6.32$\% and $R_{wp}= 8.76$\% from the neutron data.}
\label{tab:atomcoord}
\resizebox{\columnwidth}{!}{%
\begin{tabular}{ccccccc}
\hline
Atom &  Site &     $x$           &    $y$     &     $z$     & Biso(\AA$^2$)     &  Occ     \\
\hline
\hline
Pb  &  8a   &     0         &  0            &   0           &  0.76(7) &  0.5 \\
Co  &  16b  &     0.3322(13) &  0.3303(12)    &   0.1707(10)   &  0.76(7) &  1.0 \\
V   &  16b  &     0.2606(10)    &  0.0716(7)      &   0.0464(25)     &  0.76(7) &  1.0 \\
O1   &  16b  &     0.1458(5) &  0.4958(13)    &  -0.0556(9)  &  0.90(17) &  1.0 \\
O2   &  16b  &     0.3398(13) &  0.6710(15)    &   0.4273(9)  &  0.90(17) &  1.0 \\
O3   &  16b  &     0.1594(17) &  0.6816(10)    &   0.6643(10)   &  0.90(17) &  1.0 \\
O4   &  16b  &     0.3220(6) &  0.4987(11)    &   0.1415(7)   &  0.90(17) &  1.0 \\
\hline
\end{tabular}%
}
\end{table}
A combined Rietveld refinement of the X-ray and neutron diffraction pattern was performed simultaneously where all the atomic positions, the lattice parameters at 300~ and 120~K, and isotropic Debye Waller factors were included in the fit (the occupancies were constrained to be stoichiometric). 
The results of this combined Rietveld refinement (see Fig.~\ref{fig:RietvStruct}(a) and (b)) confirm the suitability of the $I4_{1}cd$ space group with lattice parameters $a=b=12.2610(12)$~{\AA}, $c= 8.3741(10)$~{\AA} found from neutron diffraction pattern at $T = 120$~K, and $a=b=12.3580(8)$~{\AA}, $c= 8.4444(5)$~{\AA} found from the X-ray diffraction pattern at room temperature. The resulting atomic position which are listed in Table~\ref{tab:atomcoord} are in good agreement with Ref.~\cite{HE20071770}. The reliability factors are $R_{F}= 6.32$\% and $R_{wp}= 8.76$\% for the neutron data and $R_{F}= 10.50$\% and $R_{wp}= 13.50$\% for the X-ray pattern. As expected the crystal structure of \pcvo is very similar to \scvo and \bcvo with the magnetic Co$^{2+}$ ions forming 4-fold screw chains running along the {\bf c}-axis, with four chains per unit cell, two rotating clockwise and two anticlockwise.

\begin{figure}[!ht]
	\centering
	\includegraphics[width=1.0\linewidth]{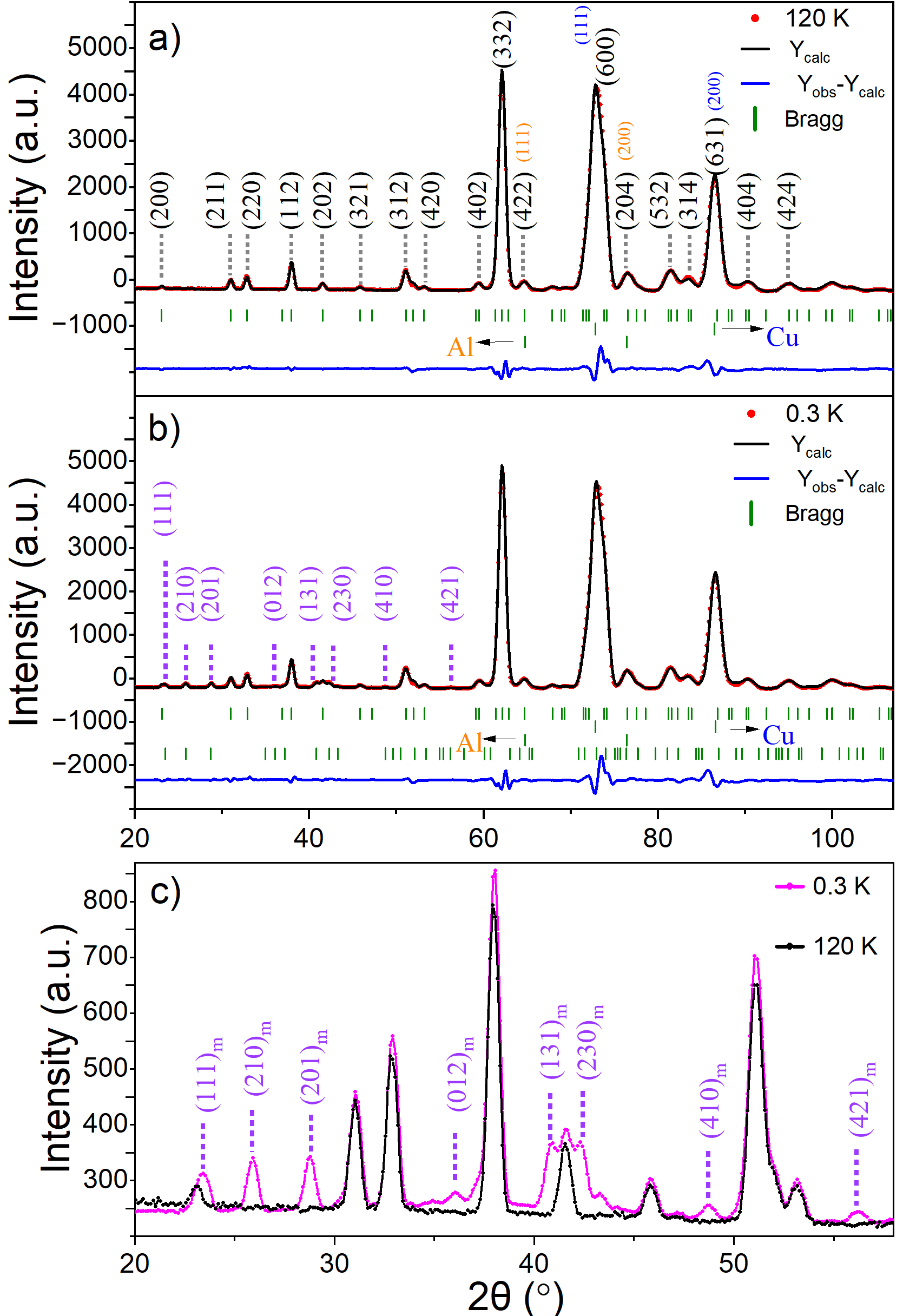}
	\caption{Neutron powder diffraction patterns from the powder sample at (a) $T = 120$~K and (b) $T = 0.3$~K (red dots). The solid black line gives the Rietveld refinement and the solid blue line gives the difference between the data and refinement. The top row of green vertical lines indicates Bragg peak positions for $I4_{1}cd$. The next two rows correspond to the Bragg peaks of Cu (sample can) and Al (shielding). The last row in (b) indicates the magnetic Bragg peaks position corresponding to the ordering wavevector $\kappa=(0,0,1)$ which are also labelled in pink. (c) The neutron diffractograms for both temperatures are overplotted for comparison in the region from 20$^{\circ}$ to 60$^{\circ}$ to highlight the magnetic peaks.}
	\label{fig:RietvMagnStruct}  
\end{figure}

For the magnetic structure determination, neutron powder diffraction patterns was collected using the \ce{^3He} insert at $T=120$~K$(\gg T_N)$ (see Fig.~\ref{fig:RietvMagnStruct}(a)) and $T=0.3$~K$(< T_{N}=3.80$~K) (see Fig.~\ref{fig:RietvMagnStruct}(b)). Additional Bragg reflections appear at integer $hkl$ positions below $T_N$ which are magnetic. The nuclear reflections occur at $h+k+l$ = 2$n$, where $n$ is an integer, due to the body-centered ($I$) symmetry. The magnetic peaks occur when $h+k+l$ = 2$n$+1, where $n$ is an integer and can be indexed by the propagation vector $\kappa=(0,0,1)$ (see, Fig.~\ref{fig:RietvMagnStruct}(c)). This is in agreement with reports on the isostructural compound \ce{SrCo2V2O8} \cite{PhysRevB.89.094402} and also with the related compound \bcvo \cite{PhysRevB.83.064421}.

\begin{table*}[btp]
\caption{Irreducible representations (IR) of the space group $I4_{1}cd$ for the propagation vector $\kappa$ = (0,0,1). The symmetry elements are written according to Wigner-Seitz notation.}
\label{tab:gamma}
\centering
\begin{tabular}{ccccccccc}
\toprule
\multirow{2}{*}{IR}                          & \multicolumn{8}{c}{Symmetry Elements} \\
 & $\{1|{000}\}$ & $\{2_{00z}|ppp\}$ & $\{4^+_{00z}|0ps\}$ & $\{4^-_{00z}|p0t\}$ & $\{m_{x0z}|00p\}$ & $\{m_{0yz}|pp0\}$ & $\{m_{x-xz}|0pt\}$ & $\{m_{xxz}|p0s\}$ \\
\hline
$\Gamma_{1}$ & 1   & -1  &  i   & -i   &  1   & -1  &  i  &  -i \\
$\Gamma_{2}$ & 1   & -1  &  i   & -i   & -1   &  1  & -i  &   i \\
$\Gamma_{3}$ & 1   & -1  & -i   &  i   &  1   & -1  & -i  &   i \\
$\Gamma_{4}$ & 1   & -1  & -i   &  i   & -1   &  1  &  i  &  -i \\
$\Gamma_{5}$ & $\begin{matrix}1& 0\\ 0& 1\end{matrix}$ & $\begin{matrix} 1& 0\\ 0& 1\end{matrix}$   & $\begin{matrix}1& 0\\ 0& -1\end{matrix}$   & $\begin{matrix}1& 0\\ 0& -1\end{matrix}$   & $\begin{matrix}0&1\\ 1&0\end{matrix}$   &  $\begin{matrix}0&1\\ 1&0\end{matrix}$  &  $\begin{matrix}0&-1\\ 1&0\end{matrix}$ &  $\begin{matrix}0&-1\\ 1&0\end{matrix}$\\
\hline
\end{tabular}%

\end{table*}

\begin{figure}[ht]
	\centering
		\includegraphics[width=1.00\linewidth]{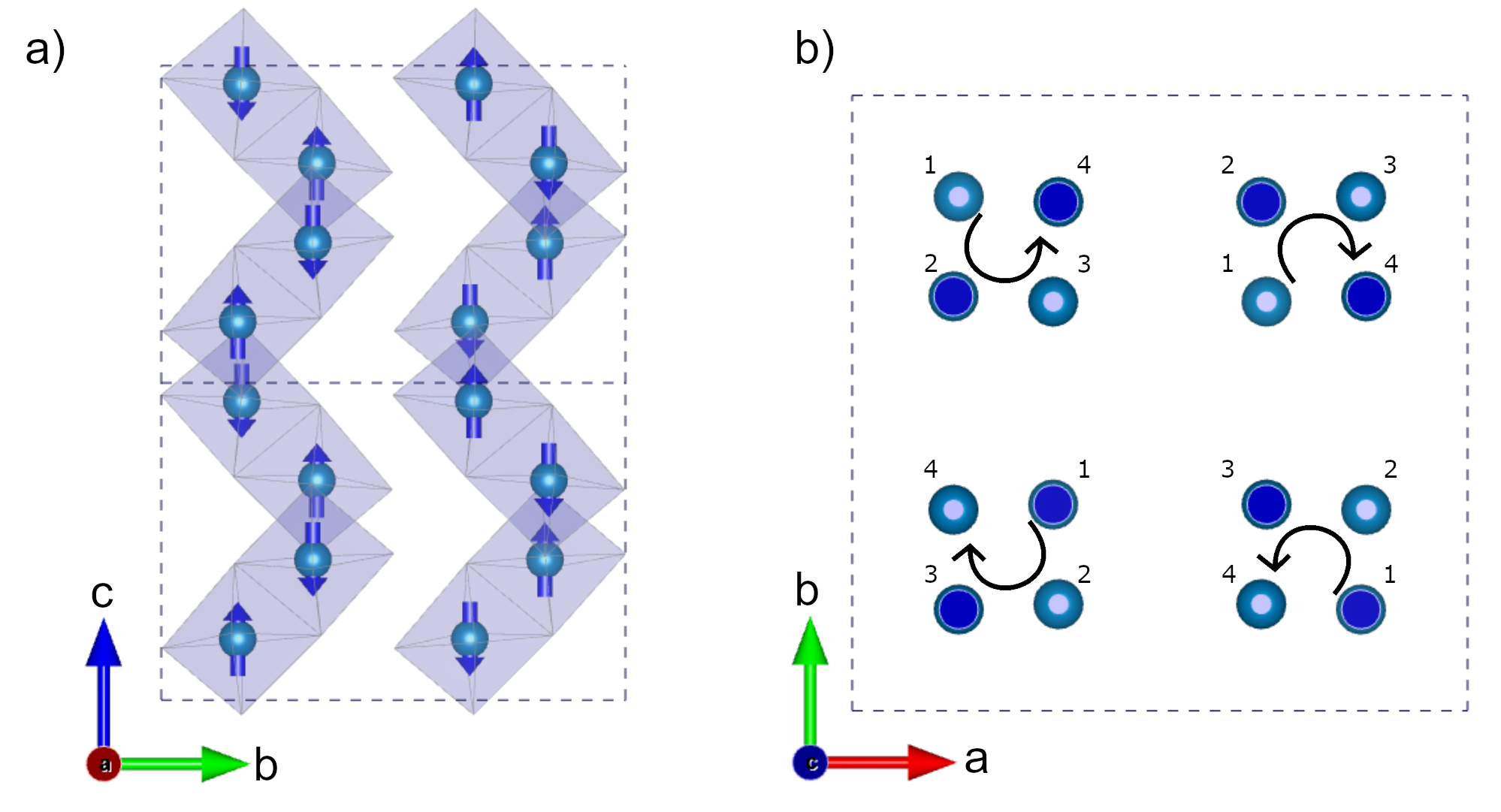}
	\caption{The magnetic structure of \ce{PbCo2V2O8} showing only the CoO$_{6}$ octahedra. (a) The projection on the {\bf b}-{\bf c}-plane shows two of the four 4-fold screw chains for clarity, the spins (blue arrows) point almost parallel to the {\bf c}-axis and are aligned antiferromagnetically along each chain. (b) The projection on the {\bf a}-{\bf b}-plane shows all four screw chains where dark blue (light blue) circles represent spin up (spin down). Black arrows show the sense of rotation of each screw chain, the Co$^{2+}$ ions of each chain are numbered 1 to 4 according to their height along the chain, thus ions with the same number are located in the same {\bf c}-plane. The spins of neighboring chains are aligned ferromagnetically (antiferromagnetically) along the {\bf a} ({\bf b}) axes respectively.}
	\label{fig:MagnStruct}
\end{figure}

To obtain the magnetic structure of \ce{PbCo2V2O8}, space group analysis was performed, using the BasiReps software from the Fullprof suite~\cite{Rodriguez1990,RODRIGUEZCARVAJAL199355}. The crystal structure is described within space group $I4_{1}cd$ (\texttt{\#} 110), which has two centering operations, and eight symmetry operations, which leave the propagation vector invariant. The magnetic representation decomposes into four one-dimensional irreducible representations (IReps) $\Gamma_{i=1-4}$ repeated three times each, and one two-dimensional IRep ($\Gamma_5$) repeated six times, \textit{i.e.} $\Gamma_{m}=3\Gamma_{1}^{1}+3\Gamma_{2}^{1}+3\Gamma_{3}^{1}+3\Gamma_{4}^{1}+6\Gamma_{5}^{2}$. The basis vectors of each irreducible representation for this space group with the propagation vector $\kappa=(0,0,1)$ and Wyckoff site ($16\,b$) for Co$^{2+}$ are presented in Table ~\ref{tab:gamma}. All possible magnetic structures were compared to the data collected at $T=0.3$~K and only $\Gamma_5$ gave good fitting parameters ($R_{F} = 17.6\%$). The resulting fit is shown in Fig.~\ref{fig:RietvMagnStruct}(b).
\par
At $T=0.3$~K the magnetic moments on the Co$^{2+}$ ions of \pcvo along the three directions are $m_a=-0.104\pm0.07$~$\mu_{B}$, $m_b=0.000$~$\mu_{B}$ and $m_c=1.436\pm0.03$~$\mu_{B}$ showing that they point predominantly along the {\bf c}-axis with a small canting away from this axis of 4$^{\circ}\pm3^{\circ}$. Within the errorbar, this canting is the same size as the canting of 3.4$^{\circ}$ of the principle (compressed) axis of the CoO$_{6}$ octahedra away from the {\bf c}-axis. The projection of this structural canting onto the {\bf a}-{\bf b} plane follows the 4-fold rotation of the screw chains. It therefore seems natural to associate the canting of the Co$^{2+}$ spins with this structural canting where the spins are constrained to point along the principle octahedral axis due to anisotropy. This was the situation found in \bcvo and \scvo where the 4-fold canting gives rise to complex g-tensor~\cite{Kimura_2013, PhysRevLett.123.067203}.

The magnitude of the ordered magnetic moment obtained from the refinement of $\mu=1.44$~$\mu_{B}$ at $T$ = 0.3 K is considerably smaller than the 3~$\mu_{B}$ expected for \ce{Co^2+} (spin only, high spin state $S=3/2$) and is also lower than the saturation magnetization of 2.75~$\mu_{B}$ we observed for magnetic field along the {\bf c}-axis. An explanation is that the Co$^{2+}$ moments are not fully ordered as is frequently observed in quasi-one-dimensional antiferromagnets where magnetic order is partially suppressed by quantum fluctuations.

Figure~\ref{fig:MagnStruct} shows the magnetic structure. The Co$^{2+}$ moments are aligned antiferromagnetically with respect to each other along each of the four 4-fold screw chains that run along the {\bf c}-axis. Figure~\ref{fig:MagnStruct}(a) gives a projection onto the {\bf b}-{\bf c}-plane, showing just two chains for clarity. 
Within the basal {\bf a}-{\bf b}-plane, The spins of neighboring chains are aligned ferromagnetically (antiferromagnetically) along the {\bf a} ({\bf b}) axes respectively (see, Fig.~\ref{fig:MagnStruct}(b)). This magnetic structure is very similar to that observed in \bcvo \cite{PhysRevB.83.064421,PhysRevB.87.054408,PhysRevB.92.134416,Faure:2017iup} and \scvo \cite{PhysRevB.89.094402,Liu2016,Shen_2019,PhysRevB.103.144405}.


\begin{figure}[h!]
	\centering
		\includegraphics[width=1.0\linewidth]{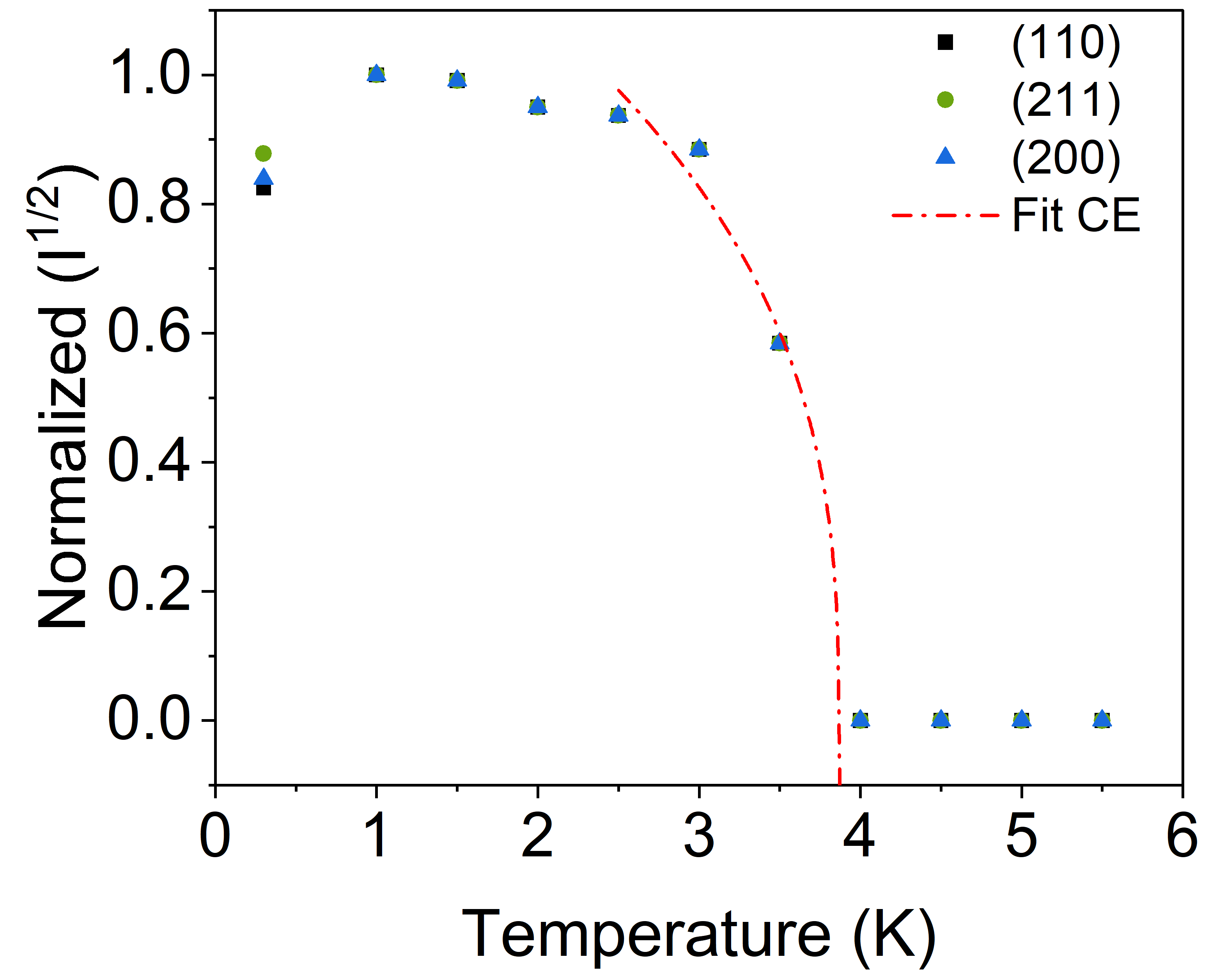}
	\caption{The temperature-dependence of the normalized square root of the integrated intensity of the $(110)$, $(21\bar{1})$, $(200)$ magnetic Bragg peaks of \ce{PbCo2V2O8}. The solid red line corresponds to the fitted curve of the critical exponent equation. }
	\label{fig:CriticExpon}
\end{figure}

One interesting point is that the space group of \bcvo ($I4_1/acd$) is centrosymmetric, while that of \scvo and \pcvo ($I4_1cd$) is non-centrosymmetric and polar. This means some ions should be displaced from their original positions in $I4_1/acd$ to off-centered position in $I4_1cd$ so as to have a polarity. Indeed the symmetry of the Co$^{2+}$ site in \scvo and \pcvo is reduced and all the Co$^{2+}$-O$^{2-}$ bonds to the surrounding octahedron are different compared to \bcvo where there are only three unique bonds. Such additional displacements should affect the magnetic properties of these materials, resulting in changes in the superexchange interactions. Despite this, the properties of all three compounds are remarkably similar in terms of their magnetic structures and phase diagrams and thus differences arising from the different space groups are very subtle.
\par
To investigate the temperature dependence of the magnetic reflections, measurements at various temperatures were performed. The integrated intensities are shown in Fig.~\ref{fig:CriticExpon} and reveal that the magnetic intensity disappears above 4~K due to the loss of long-range magnetic order. The data were compared to the power law function $I=I_{0}+A(1-\frac{T}{T_N})^{2\beta}$ where $A$ is a proportionality constant. Magnetic systems are classified into several categories with different values of temperature exponent $\beta$, such as $\beta$ = 0.326 (3D Ising), $\beta$ = 0.35 (3D XY), $\beta$ = 0.367 (3D Heisenberg), and $\beta$ = 0.50 (mean-field) \cite{PhysRevMaterials.7.014402,blundell2001magnetism}. For \ce{PbCo2V2O8}, this equation was fitted to the data over the temperature range of 2.5 K to 3.5 K, keeping the critical temperature fixed at $T_N=3.80$~K which was the value obtained from our heat capacity (Section \ref{sec:HC}) and magnetization (Section \ref{sec:TdepMag}) measurements. The fit yielded the exponent $\beta=0.316\pm0.06$. This value of $\beta$ might be an indication for the 3D Ising model, although the limited amount of data significantly reduces the reliability of this result. For comparison the value $\beta=0.33\pm0.03$ was found for \scvo \cite{PhysRevB.89.094402} and $\beta=0.307-0.328$ \cite{PhysRevB.87.054408} or $\beta=0.28$ \cite{PhysRevB.83.064421} for \bcvo. 


\section{Summary}


To conclude we have synthesized powder and, to our knowledge, the first single crystal samples of PbCo$_2$V$_2$O$_8$, which we have investigated using magnetization, specific heat, and neutron diffraction. The crystal structure of \pcvo gives rise to 4-fold screw chains of magnetic Co$^{2+}$ ions along the {\bf c}-axis. In zero magnetic field, long-range magnetic order takes place at $T_N=3.80$~K and the moments order antiferromagnetically along the chains with the spins canted a small amount from the {\bf c}-axis. We confirm the presence of a Heisenberg-Ising (XXZ-type) anisotropy and application of a magnetic field gives rise to a complex series of new phases that are different for the $H\|${\bf c}, $H\|${\bf a} and $H\|$[110] directions. We have constructed detailed phase diagrams for all three directions up to 11.5~T and have also explored the behavior to high fields. Apart from having a slightly smaller energy scale, \pcvo shows many similarities to \scvo and \bcvo. The resemblance of their phase diagrams for $H\|${\bf c} and $H\|${\bf a} suggests that the same magnetic phases occur in all three compounds for these directions. However, the phase diagram for the $H\|$[110] field direction reveals a new phase for \pcvo at $H\|$[110]~$\approx 10$~T, which has not been previously reported in either \scvo or BaCo$_2$V$_2$O$_8$. The nature of this phase is unknown and it is not clear why it appears in \pcvo but not the other compounds. One possibility is that the phase is in fact present in \scvo or \bcvo but their higher energy scale drives it to higher fields making it inaccessible due to the low temperature necessary for its observation which are unavailable in high field magnets, and thus it has been missed until now.

As model spin chain materials with XXZ anisotropy and complex g-tensors, the ACo$_2$V$_2$O$_8$ A = Sr, Ba compounds have remarkably rich phenomena and have been used to test and explore several different fundamental physics ideas. Finding a new member of this family raises the possibility of finding new phenomena and further refining theories. To this end, we plan to investigate the origin of the new phase for $H\|$[110]~$> 10$~T in \pcvo starting with single crystal neutron diffraction in the near future. \\

\begin{acknowledgments}


We acknowledge the Core Lab Quantum Materials
(CLQM), Helmholtz Zentrum Berlin für Materialien und
Energie (HZB), Germany, where the powder and single crystal samples of PbCo$_2$V$_2$O$_8$ were synthesized and measured. The authors acknowledge the support of Hochfeld Magnetlabor Dresden at Helmholtz Zentrum Dresden Rossendorf (HLD-HZDR), a member of the European Magnetic Field Laboratory (EMFL). This work is also partly based on experiments performed at the Swiss spallation neutron source SINQ, Paul Scherrer Institute, Villigen, Switzerland. 
\end{acknowledgments}
\typeout{}
\bibliography{cite}

\bibliographystyle{apsrev4-2}

\end{document}